\documentclass[12pt]{article}
\usepackage{graphicx}

\newcommand{\beq}{\begin{equation}}
\newcommand{\eeq}{\end{equation}}
\newcommand{\beqn}{\begin{eqnarray}}
\newcommand{\eeqn}{\end{eqnarray}}
\newcommand{\beqns}{\begin{eqnarray*}}
\newcommand{\eeqns}{\end{eqnarray*}}

\begin{document}

\begin{titlepage}
\begin{center}

\hfill USTC-ICTS/PCFT-22-05\\
\hfill February 2022

\vspace{2.5cm}

{\large {\bf Higgs boson decays into a pair of heavy vector quarkonia}}\\
\vspace*{1.0cm}
 {Dao-Neng Gao$^\dagger$ and Xi Gong$^\ddagger$\vspace*{0.3cm} \\
{\it\small Interdisciplinary Center for Theoretical Study,
University of Science and Technology of China, Hefei, Anhui 230026
China}\\
{\it\small Peng Huanwu Center for Fundamental Theory, Hefei, Anhui 230026 China}}

\vspace*{1cm}
\end{center}
\begin{abstract}
\noindent
Rare Higgs decays into a pair of heavy vector quarkonia, $h\to VV$ ($V=J/\Psi$, $\Upsilon$ etc.), have been investigated in the standard model. Different from the past literature in which these decays are thought to be only dominated by the longitudinally polarized final states, we also include the transitions, which proceed through
$h\to \gamma^*\gamma^*$/$h\to V\gamma^*$, followed by $\gamma^*\to V$. The final vector quarkonia via these ways are dominantly transversely polarized. Our calculation however shows that these transitions could lead to significant contributions to the decay rate, especially for the charmonium final states. The total branching ratios of these processes are predicted to be around $10^{-10}$, far below the current experimental upper bounds. Hopefully, experimental studies of these very rare decays in future high-precision experimental facilities might be interesting both to test the standard model and to look for new physics scenarios.
\end{abstract}

\vfill
\noindent
$^{\dagger}$ E-mail address:~gaodn@ustc.edu.cn\\\noindent
$^{\ddagger}$ E-mail address:~gonff@mail.ustc.edu.cn
\end{titlepage}

 After the discovery of the Higgs boson at the Large Hadron Collider (LHC) \cite{atlascms}, the focus has shifted to the precise determination of the properties of this newly discovered particle. Comprehensive experimental studies indicate that the measured couplings of the Higgs boson to standard model (SM) fields \cite{atlascmsjoint, cmsatlas19} are compatible, so far, with their SM values.  On the other hand, exclusive rare Higgs decays, which might be very interesting at the future high-energy and/or high-luminosity
 experimental facilities, have been investigated both theoretically and experimentally, such as $h\to V \gamma$ \cite{Wilczek77, BS79, Keung83, BPSV13, KPPSSZ14, KN15, atlas15}, $h\to V Z$ \cite{IMT13, Gao,BAL14,MS14,AKN16,Zhao18}, and $h\to V \ell \bar{\ell}$ \cite{CFS16,PS16, GG21} decays, with $V$ denoting vector mesons $\rho$, $\phi$, $J/\psi$ and $\Upsilon$ etc. Due to their small SM branching ratios, it will be in general challenging to search for these rare processes, however, experimental studies of them may be helpful both to increase our understanding of the properties of SM Higgs boson and to potentially probe the novel Higgs dynamics in new physics scenarios.

Recently, a search for rare Higgs decays into a pair of heavy vector quarkonia, $h \to V V $ ($V=J/\Psi, \Upsilon$), has been firstly performed by the CMS Collaboration \cite{CMS19}, and upper limits on the branching fractions have been measured to be
\beq\label{brJPsi}
{\cal B}(h\to J/\Psi J/\Psi) < 1.8\times 10^{-3}
\eeq
and
\beq\label{brUpsilon}
{\cal B}(h\to \Upsilon(1S) \Upsilon(1S)) < 1.4\times 10^{-3}
\eeq
at the 95\% confidence level, respectively.

Theoretically, these processes have already been studied in the literature \cite{BS79, Keung83, KLN09}. In particular, the authors of Ref. \cite{KLN09} have explicitly calculated the decay rates for $h\to V V$ in the SM, by assuming that the dominant contributions should be from the transitions in which final state vector quarkonia are  longitudinally polarized. The typical diagrams are shown in Figure 1.
Theoretical predictions, depending on the Higgs boson mass, for the branching ratios of these decays have been given in their paper \cite{KLN09}, and one can easily read that, as stated in Ref. \cite{CMS19}, the values are about
\beq\label{brJPsi-th1}
{\cal B}(h\to J/\Psi J/\Psi)=1.5\times 10^{-10}
\eeq
and
\beq\label{brUpsilon-th1}
{\cal B}(h\to \Upsilon(1S) \Upsilon(1S))=2\times 10^{-9}
\eeq
for $m_h=125$ GeV.

The purpose of the present paper is to reexamine the analysis of $h\to V V$ decays in the SM. It will be shown below that, besides the diagrams in Figure 1, some other diagrams, as displayed in Figure 2, cannot be neglected for these transitions. Thus the Higgs boson decays into a pair of vector heavy quarkonia could also occur via $h\to \gamma\gamma$ or $h\to  V \gamma$, with the virtual photon transforming into $V$. These diagrams may lead to significant contributions to the decay rate, even larger than those from Figure 1 for charmonium final states. Similar mechanism has been studied in $h\to \gamma V$ \cite{BPSV13} and $h\to Z V$ \cite{Gao} decays.  Therefore, it is of interests to perform a systematical calculation of the branching fractions of $h\to VV$ decays,  including all of the relevant diagrams,  in order that one can compare the SM predictions with the future experimental measurements.

\begin{figure}[t]
\begin{center}
\includegraphics[width=12cm,height=3.0cm]{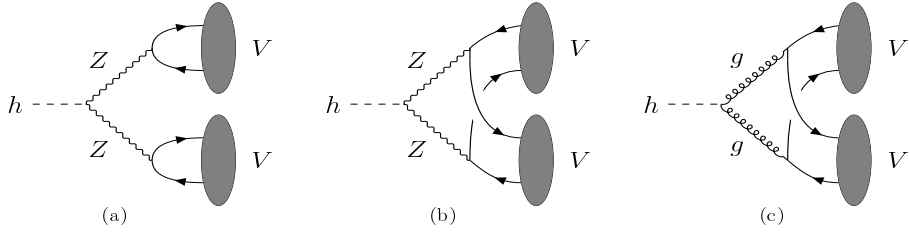}
\end{center}
\caption{Diagrams contributing to $h\to VV$ decays in the SM: (a) and (b) through the $ZZ$ intermediate state while (c) through the gluon intermediate state. The solid line with arrow denotes the heavy quarks $Q$ or $\bar{Q}$.}\label{figure1}
\end{figure}

First let us evaluate the decay amplitudes from Figure 1, which can be split as
 \beq\label{amp1}
 {\cal M}_1={\cal M}_{1(a)}+{\cal M}_{1(b)}+ {\cal M}_{1(c)}.
 \eeq
 One can easily find that these diagrams contain the couplings of the Higgs boson to a pair of $Z$ bosons or gluons, which are in turn converted to two heavy quark pairs $Q\bar{Q}$. To further obtain the hadronic decay amplitudes, one has to project $Q\bar{Q}$ into the corresponding hadron states. Here we will adopt the nonrelativistic color-singlet model \cite{colorsinglet}, as a reasonable approximation for the leading order calculation. Within this model the quark momentum and mass are taken to be one half of the corresponding quarkonium momentum $p$ and mass $m_V$, which means $p_Q=p_{\bar{Q}}=p/2$ and $m_{V}=2m_Q$.  Thus for the $Q\bar{Q}$ pair to form the heavy quarkonium $V$, according to Refs. \cite{barger, Jia2007}, one can replace the combination of the Dirac spinors for $Q$ and $\bar{Q}$ by the following projection operator
\beq\label{projector}
{v(p_{\bar{Q}})} {\bar{u}(p_Q)} \longrightarrow \frac{\psi_{V}(0)I_c}{2\sqrt{3m_V}} {\epsilon\!\!/}^*(p\!\!\!/+m_V),
\eeq
where $I_c$ is the $3\times 3$ unit matrix in color space and $\epsilon^{*\mu}$ is the polarization vector of the heavy quarkonium $V$. $\psi_{V}(0)$ is the wave function at the origin for $V$, which is a nonperturbative parameter.

Now using the standard vertices of $hZZ$ and $Z Q\bar{Q}$, one can straightforwardly derive the decay amplitudes of $h\to V V$ from Figure 1(a) and 1(b), which read
\beqn\label{amp1(a)}
&&{\cal M}_{1(a)}=\frac{6(g_v^Q)^2 g^2}{v \cos^2\theta_W}\frac{m_Z^2}{(m_Z^2-m_V^2)^2} \left(\frac{\psi_V(0)}{\sqrt{m_V}}\right)^2 m_V^2 \epsilon^*(p)\cdot \epsilon^*(q),
\\\nonumber\\
\label{amp1(b)}
&&{\cal M}_{1(b)}=\frac{[(g_v^Q)^2+(g_a^Q)^2]g^2}{v \cos^2\theta_W}\frac{m_Z^2}{(m_Z^2-m_h^2/4)^2} \left(\frac{\psi_V(0)}{\sqrt{m_V}}\right)^2 m_V^2 \epsilon^*(p)\cdot \epsilon^*(q).
\eeqn
Here $p$ and $q$ represent the momentum of the two vector quarkonia in the final states, respectively. $v=(\sqrt{2}G_F)^{-1/2}\approx 246$ GeV,  $g$ is the weak SU(2)$_L$ coupling constant, and $\theta_W$ is the Weinberg angle. $g_v^Q=T_3^Q-2 Q_q \sin^2 \theta_W$ and $g_a^Q=T_3^Q$,  where $Q_q$ is the charge and $T_3^Q$ is the third component of the weak isospin of the heavy quark.

For the case of the gluon intermediate state in Figure 1(c), it is known that the Higgs coupling to gluons is absent at the tree level in the SM,  and nonzero contributions to the effective $h g g $ vertex is mediated by heavy quark loop, with top quark providing the dominant one. Explicitly, one can write down the effective lagrangian, at the lowest order, for this coupling as follows \cite{effective-hgg}
\beq\label{effective-hgg}
{\cal L}_{h gg}=\frac{\alpha_s}{12\pi v} G^{a}_{\mu\nu}G^{a\mu\nu} h
\eeq
with $\alpha_s$ denoting the strong coupling constant and $G^{a}_{\mu\nu}$ the field-strength tensor for gluons. Thus direct calculations will give \footnote{In general, a momentum-dependent coefficient in eq. (\ref{effective-hgg}) should be adopted to include the effects from the virtual gluons. However, we have found, by explicit calculation, that this correction to eq. (\ref{amp1(c)}) is below $2\%$, which can be safely neglected.}
\beq\label{amp1(c)}
{\cal M}_{1(c)}=-\frac{128 \alpha_s^2}{9 v m_h^2}\left(\frac{\psi_V(0)}{\sqrt{m_V}}\right)^2 m_V^2 \epsilon^*(p)\cdot \epsilon^*(q).
\eeq

\begin{figure}[t]
\begin{center}
\includegraphics[width=12cm,height=3.0cm]{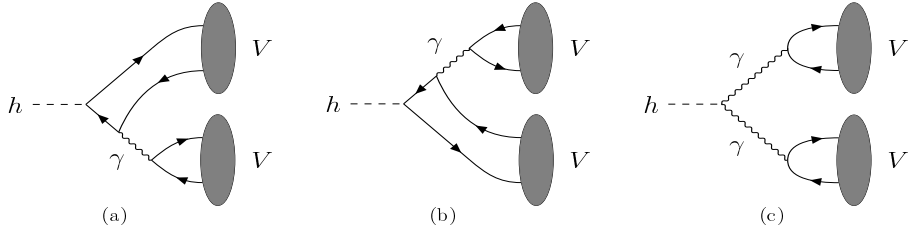}
\end{center}
\caption{Diagrams contributing to $h\to VV$ decays in the SM through the virtual photon intermediate states. The solid line with arrow denotes the heavy quarks $Q$ or $\bar{Q}$.  }\label{figure2}
\end{figure}

Similarly, one can deal with Figure 2,  and in these diagrams the $h\to V V$ transitions proceed through the $h\gamma\gamma$ and $h\to V\gamma$ (via the $h Q\bar{Q}$ coupling), respectively. Note that the leading-order $h\gamma\gamma$ interaction in the SM is induced by one-loop diagrams involving $W$-boson and heavy charged fermions like top-quark. Effectively, one may write
\beq\label{effectivehgammagamma}
{\cal L}_{h\gamma \gamma}=\frac{e^2}{32\pi^2 v} C_{\gamma} F_{\mu \nu} F^{\mu\nu} h,
\eeq
where $e$ is the QED coupling constant and the SM value of the dimensionless efficient $C_\gamma$ can be found in Ref. \cite{BH85}.  Therefore, we have
\beq\label{amp2}
{\cal M}_2={\cal M}_{2(a)+2(b)}+{\cal M}_{2(c)}
\eeq
with
\beq\label{amp2(a)+(b)}
{\cal M}_{2(a)+2(b)}=\frac{48 Q_q^2 e^2}{v m_h^2}\frac{1}{m_V^2}\left(\frac{\psi_V(0)}{\sqrt{m_V}}\right)^2 m_V^2 (p^\mu q^\nu-p\cdot q g^{\mu\nu})\epsilon^*_\nu (p) \epsilon^*_\mu (q),
\eeq
and
\beq\label{amp2(c)}
{\cal M}_{2(c)}=\frac{3 Q_q^2 e^4 C_\gamma}{2\pi^2v}\frac{1}{(m_V^2)^2}\left(\frac{\psi_V(0)}{\sqrt{m_V}}\right)^2 m_V^2 (p^\mu q^\nu-p\cdot q g^{\mu\nu})\epsilon^*_\nu (p) \epsilon^*_\mu (q).
\eeq
We keep the factor $1/m_V^2$ in above equations in order to show its origin from the virtual photon propagator. Here we should emphasis that, following the treatment in Refs. \cite{BBLY06, BPSV13}, Figure 2(c) can also be calculated by observing that the virtual photon couples to a vector quarkonium through a matrix element of the electromagnetic current. This approach has the advantage that it automatically takes into account higher order corrections to the electromagnetic current,which are common to both the electromagnetic decay and production of a vector meson.

\begin{figure}[t]
\begin{center}
\includegraphics[width=6cm,height=3cm]{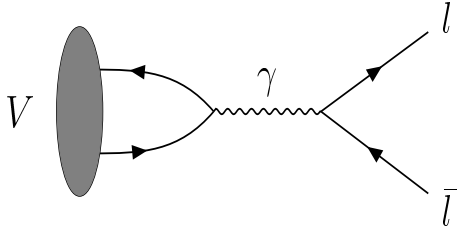}
\end{center}
\caption{Lowest-order diagram for the decay $V \to l^+ l^-$.}\label{figure3}
\end{figure}

Thus the total decay amplitude of $h\to VV$ is
\beq\label{totamplitude}
{\cal M}={\cal M}_1+{\cal M}_2,\eeq
and the decay rate will be obtained by squaring the amplitude ${\cal M}$ and summing over the polarizations of final particles, which can be expressed as
\beq\label{total-rate}
\Gamma(h\to V V)=\Gamma_1+\Gamma_2+\Gamma_{12}.\eeq
Here $\Gamma_i$ denotes the contribution from ${\cal M}_i$ for $i=1,2$, respectively, and $\Gamma_{12}$ is given by the interference between ${\cal M}_1$ and ${\cal M}_2$.
In terms of the total Higgs decay width $\Gamma_h$, we define
\beq\label{branchingratioi}
{\cal B}_i(h\to V V)={\Gamma_i}/{\Gamma_h}
\eeq
for $i=1,2$, and
\beq\label{branchingratio}
{\cal B}(h\to V V)=\frac{\Gamma(h\to V V)}{\Gamma_h}.
\eeq

To further illustrate the numerical results, we need the value of $\psi_V(0)$, which could be reached from the experimental partial width of the quarkonium decay to a pair of the charged leptons. The lowest-order contribution to this transition, as depicted in Figure 3, can be given by
\beq\label{Vtoll}
\Gamma(V\to l^+ l^-)=\frac{16 \pi \alpha_{\rm em}^2 Q_q^2}{m_V^2}\left|\frac{\psi_V(0)}{\sqrt{m_V}}\right|^2,
\eeq
where $\alpha_{\rm em}=e^2/4 \pi$. Now by taking the experimental data of $\Gamma(V\to e^+ e^-)$ from Ref. \cite{PDG},  one can predict branching ratios of $h\to VV$ decays. Numerical results have been summarized in Table 1, and the theoretical value for the total SM Higgs width, $\Gamma_h=4.10$ MeV, referring to $m_h=125.09$ GeV \cite{LHCHiggsgroup}, has been used in the calculation. In the present paper, we perform our study in the framework of the nonrelativistic color-singlet model. In order to improve our predictions, one may use the nonrelativistic QCD (NRQCD) factorization method \cite{BBL95} to calculate corrections in powers of $\alpha_s$ and $v$, where $v$ is the heavy-quark velocity in the quarkonium rest frame. Our results will be equal to the ones from the NRQCD approach at the leading order. To estimate error bars on the numerical results in Table 1, we have simply assumed that the uncalculated QCD corrections in $\alpha_s$ are of relative size $\alpha_s(m_V)$
and that the uncalculated corrections in $v$ are of relative size $v^2$, by taking $\alpha_s(m_V)\approx 0.25$ and $v^2 \approx 0.3$  for charmonium, and $\alpha_s(m_V)\approx 0.18$ and $v^2\approx 0.1$ for bottomonium, respectively. Certainly, a systematical analysis of uncertainties of our predictions from higher order $\alpha_s$ and $v$ in the framework of NRQCD would be an interesting topic for the future study.

\begin{table}[t]\begin{center}\begin{tabular}{ c  c  c c c} \hline\hline
 $V$ & $m_{V}$(GeV)& ${\cal B}_1(h\to V V)$ &${\cal B}_2(h\to V V)$ & ${\cal B}(h\to V V)$\\\hline
 $J/\Psi(1S)$ & 3.097&$(1.6\pm 0.6)\times 10^{-11}$ &$(5.8\pm 2.2)\times 10^{-10}$ & $(5.9\pm 2.3)\times 10^{-10}$\\
$\Psi(2S)$& 3.686 &$(3.9\pm 1.5)\times 10^{-12}$ &$(4.7\pm 1.8)\times 10^{-11}$ &$(5.1\pm 2.0)\times 10^{-11}$\\
$\Upsilon(1S)$& 9.460&  $(4.1\pm 0.8)\times 10^{-10}$& $(2.4\pm 0.5)\times 10^{-11}$ & $(4.3\pm 0.9)\times 10^{-10}$\\
$\Upsilon(2S)$& 10.02&  $(9.6\pm 1.9)\times 10^{-11}$&$(6.4\pm 1.3)\times 10^{-12}$& $(1.0\pm 0.2)\times 10^{-10}$\\
$\Upsilon(3S)$&10.36& $(5.4\pm 1.1)\times 10^{-11}$&$(3.8\pm 0.8)\times 10^{-12}$&$(5.7\pm 1.2)\times 10^{-11}$\\\hline
\hline
\end{tabular}\caption{Branching ratios of $h\to V V$ decays with $V$ denoting the narrow $c\bar{c}$ and $b\bar{b}$ heavy vector quarkonia.} \end{center}\end{table}

It is obvious that our results do not agree well with the ones [eqs.(\ref{brJPsi-th1}) and (\ref{brUpsilon-th1})] given in Ref. \cite{KLN09}. As mentioned above, these authors only calculated the diagrams in Figure 1, and assumed that, at the leading order, it would suffice to take into account the longitudinally polarized final state vector quarkonia in these transitions. This claim is on the basis that, for the energetic vector meson with the momentum $p$, its longitudinal polarization could be as $\epsilon^{*\mu}_L=p^\mu/m_V$, up to corrections of second order in $m_V/m_h$. By comparison, contributions from other components like transversely polarized ones will be suppressed by a factor $\sim {\cal O}(m_V/m_h)$, which thus in general can be neglected in the case of $m_V\ll m_h$. Using these arguments, one can easily show that, for the amplitude ${\cal M}_1$ from Figure 1, its polarization structure $\epsilon^*(p)\cdot \epsilon^*(q)$ [see eqs.(\ref{amp1(a)}), (\ref{amp1(b)}), and (\ref{amp1(c)})] will predominantly be $\epsilon^*_L(p)\cdot \epsilon^*_L(q)\to p\cdot q/m_V^2$, and give rise to a factor $\sim {\cal O}(m_h^2/m_V^2)$; while for the amplitude ${\cal M}_2$ from Figure 2,  its polarization structure looks like $(p^\mu q^\nu-p\cdot q g^{\mu\nu})\epsilon^*_\nu (p) \epsilon^*_\mu (q)$ [see eqs. (\ref{amp2(a)+(b)}) and (\ref{amp2(c)})],
the replacement of $\epsilon^{*\mu}\to p^\mu/m_V$ will lead to the vanishing results. This indicates that ${\cal M}_2$ should be dominated by the transversely polarized final states instead of the longitudinally polarized ones, and it seems that its contributions might be strongly suppressed and negligible. However, note that the virtual photon propagator appearing in Figure 2 will give an extra factor $1/m_V^2$ via $\gamma^*\to V$, therefore ${\cal M}_2$ could be at the same order as ${\cal M}_1$ for small $m_V$. Consequently, these diagrams will also significantly contribute to Higgs decays into a pair of heavy vector quarkonia.

Actually, the lowest-order QCD contribution to $h\to VV$ decays, as displayed in Figure 4, proceeds via the tree-level vertex $h\to Q\bar{Q}$, which was first calculated by the author of Ref. \cite{Keung83}. It has been shown in \cite{Keung83} that this amplitude has the same polarization structure as that for ${\cal M}_2$. However, different from the case of Figure 2, now we cannot expect any extra $1/m_V^2$ factor. This implies that these diagrams will be suppressed by $m_V^2/m_h^2$. Therefore we do not include them in our calculation. This suppression has also been pointed out by the authors of Ref. \cite{KLN09}.

\begin{figure}[t]
\begin{center}
\includegraphics[width=8cm,height=3.0cm]{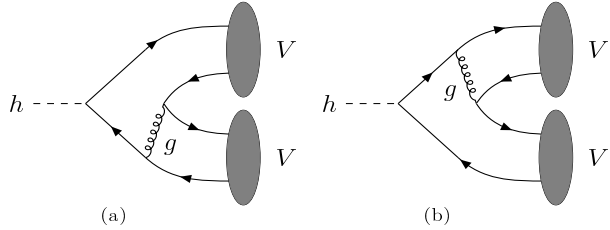}
\end{center}
\caption{Lowest-order QCD Diagrams contributing to $h\to VV$ decays. The solid line with arrow denotes the heavy quarks $Q$ or $\bar{Q}$.  }\label{figure4}
\end{figure}

On the other hand, there may exist the diagram like Figure 1(c) in which gluons are replaced by photons. This diagram has the same powers of the electromagnetic coupling as Figure 2(c). However, due to the virtual photon propagators in Figure 2(c) giving rise to the factor $(1/m_V^2)^2$, contributions from this diagram will be suppressed by a factor $m_V^2/m_h^2$, relative to eq. (14). We thus neglect it in the present calculation.

The main results of this work are given in Table 1. Our predictions for ${\cal B}(h\to VV)$ in the SM are around $10^{-10}$, far below the present experimental upper bounds given by the CMS Collaboration \cite{CMS19}. Numerically, for the charmonium $J/\Psi$ and $\Psi(2S)$ final states, contributions from Figure 2 are very significant; while for the bottomonium ones, Figure 1 gives the dominant contribution. By looking carefully at the amplitude ${\cal M}_2$, which consists of two parts, ${\cal M}_{2(a)+2(b)}$ and ${\cal M}_{2(c)}$, we have
\beq\label{amp2-1}
{\cal M}_2={\cal N}\frac{Q_q^2 e^2}{v m_h^2}\left(\frac{\psi_V(0)}{\sqrt{m_V}}\right)^2 (p^\mu q^\nu-p\cdot q g^{\mu\nu})\epsilon^*_\nu (p) \epsilon^*_\mu (q),
\eeq
where
\beq\label{Ntilde}
{\cal N}=48+\frac{6 \alpha_{\rm em} C_\gamma}{\pi}\frac{m_h^2}{m_V^2}
\eeq
is a dimensionless constant. The first factor in eq. (\ref{Ntilde}) is contributed by Figure 2(a) and 2(b), and the second one is by Figure 2(c). Explicitly, for example,  ${\cal N}\simeq -110$ for $m_V=m_{J/\Psi}$ and ${\cal N}\simeq 31$ for $m_V=m_{\Upsilon(1S)}$. Although the cancellation could happen in ${\cal N}$ due to the negative value of $C_\gamma$, the small masses of $J/\Psi$ and $\Psi(2S)$ gives rise to a large factor $m_h^2/m_V^2$,  thus ${\cal B}_2$  will be rather enhanced for charmonium quarkonia final states. However, different from the $c\bar{c}$ case, the relative large masses of $b\bar{b}$ states and the electric charge of bottom quark ($Q_b=-1/3$, $Q_c=2/3$) will lead to a suppression factor in ${\cal B}_2$. Similar situations have also occurred in $h\to \gamma V $ \cite{BPSV13} and $h\to ZV$ decays \cite{Gao}.

Furthermore, from Table 1, even if only focusing on Figure 1, our numerical results are still not consistent with predictions given in Ref. \cite{KLN09}. In the present paper, for heavy vector quarkonia, we adopt the nonrelativistic color-singlet model to perform the calculation; while the light-cone distribution amplitude approach was used by the authors of Ref. \cite{KLN09}. Therefore it is not very easy to directly compare these two results. On the other hand, the amplitude ${\cal M}_{1(c)}$ from Figure 1(c) should be ${\cal O}(\alpha_s^2)$,  since the effective $hgg$ vertex is already ${\cal O}(\alpha_s)$. However, the amplitude from the same diagram was given as ${\cal O}(\alpha_s)$ in eq. (12) of Ref. \cite{KLN09}, which might enhance their numerical results. It should be interesting to carry out a systematical calculation of these diagrams for $h\to VV$ decays using the light-cone distribution amplitude approach. This will be left for future work.

To summarize, we have presented a theoretical study of Higgs decays into a pair of heavy vector quarkonia in the SM. It was assumed in the past literature that $h\to VV$ decays should be dominated by the longitudinally polarized final states, and other contributions from the transversely polarized components would be suppressed by $m_V/m_h$. Our study indicates that this is not the whole story. Actually, these decays can also proceed through $h\to \gamma^*\gamma^*$/$h\to V\gamma^*$, with the subsequent transition $\gamma^*\to V$. Although final vector quarkonia in these cases are dominantly transversely polarized, the appearance of the virtual photon propagator due to $\gamma^*\to V$ will give rise to the factor $1/m_V^2$, which thus counteracts the above suppression. We include all relevant diagrams in our calculation. Numerical analysis shows that, for the charmonium case, transversely polarized final states give the dominant contribution; while the longitudinally polarized ones are more important for the bottomonium modes.

The total decay rates of $h\to VV$ in the SM have also been calculated, and our predictions for their branching fractions are around $10^{-10}$, which are far from the present experimental limits. To search for these very rare decays will be very challenging experimentally. On the other hand, this may indicate that substantial room for new physics could be expected in these processes. Therefore, in the future high-precision experiments, it is interesting to explore these decays both to increase our understanding of the SM and to probe new physics beyond the SM.

\vspace{0.5cm}
\section*{Acknowledgements}
We are very grateful to the anonymous referee for constructive comments which help to improve the present paper. This work was supported in part by the National Natural Science Foundation of China under Grants No. 11575175, No. 11947301, and No. 12047502, and by National Research and development Program of China under Contract No. 2020YFA0406400.

\end{document}